\documentclass[prc,aps,twocolumn,floats,floatfix,showpacs,superscriptaddress,nofootinbib]{revtex4}

\usepackage{amsmath}
\usepackage{amsfonts}
\usepackage{graphicx}

\newcommand{\nuc}[2]{$^{#1}${#2}}
\newcommand{\etal}{\emph{et al.}}

\begin{document}

\title{Large-amplitude $Q_n$-$Q_p$ collectivity in the neutron-rich
       oxygen isotope \nuc{20}{O}}

\author{A. P. Severyukhin}
\affiliation{Bogoliubov Laboratory of Theoretical Physics,
             Joint Institute for Nuclear Research,
             141980 Dubna, Moscow region, Russia}

\author{M. Bender}
\affiliation{Dapnia/SPhN, CEA Saclay,
             F-91191 Gif sur Yvette Cedex,
             France}
\affiliation{Universit{\'e} Bordeaux 1; CNRS/IN2P3;
             Centre d'Etudes Nucl{\'e}aires de Bordeaux Gradignan, UMR5797,
             Chemin du Solarium, BP120, F-33175 Gradignan, France}

\author{H. Flocard}
\affiliation{CNRS-IN2P3, Universit\'e Paris XI, CSNSM, Bt. 104,
             F-91405 Orsay Campus, France}

\author{P.-H. Heenen}
\affiliation{PNTPM, CP229,
             Universit{\'e} Libre de Bruxelles,
             B-1050 Brussels, Belgium}

\begin{abstract}
By means of HFB calculations with independent constraints on axial
neutron and proton quadrupole moments $\hat{Q}_n$ and $\hat{Q}_p$,
we investigate the large amplitude isoscalar and isovector
deformation properties of the neutron-rich isotope \nuc{20}{O}.
Using the particle-number and angular-momentum projected Generator
Coordinate Method, we analyze the collective dynamics in the
$\{\langle \hat{Q}_n\rangle\, ,\langle \hat{Q}_p\rangle \}$ plane.
The parameterization SLy4 of the Skyrme interaction is used for
all calculations in connection with a density-dependent zero-range
pairing interaction. Our results show that already for this
moderately neutron-rich nucleus the transition moments are modified 
when independent neutron and proton collective dynamics are allowed.
\end{abstract}


\date{January 12 2007}

\maketitle
%
%
\section{Introduction}

The dominance of the neutron-proton attraction is one of the main
characteristics of the effective nucleon-nucleon interaction. It
can be traced back to the isospin symmetry of the force and to the
cooperation of the \mbox{$T=0$} and \mbox{$T=1$} channels. As a
result, isovector collective modes have a much higher excitation
energy than their isoscalar counterparts. On the other hand, the
study of nuclei away from the stability line has led to the
discovery of phenomena (halo and molecular nuclei, reorganization
of the shell structure) which challenge some of the standard
nuclear structure concepts. These new perspectives have been
mostly provided by the analysis of neutron-rich nuclei which,
because of the relative weakening of Coulomb effects, give the
best opportunities to investigate dynamics far away from the
\mbox{$N=Z$} line. The new phenomena are partly attributed to a
decoupling of the two nucleonic distributions when $N$ exceeds $Z$
by far. Another cause invoked is the weak binding of the last
occupied neutrons in the absence of a barrier. Aside from new
features of ground-state properties, also some interesting
effects, such as pigmy resonances, have been observed which are
clear indications that the excitation of isovector modes is easier
in neutron-rich isotopes than in stable nuclei. The pigmy modes
are associated with dipolar excitations involving a restricted
number of nucleons and partly for this reason, carrying a small
fraction of the sum rule only. They also correspond to collective
vibrations. On the other hand, most of the low-energy collective
structure remains dominated by isoscalar large amplitude
quadrupole dynamics, leading in many regions of the nuclear chart
to ground state deformations.

The strong proton shell closure at \mbox{$Z=8$} and the already good
experimental knowledge of the complete oxygen isotope series from
\nuc{16}{O} up to the drip line at \nuc{24}{O} \cite{Ram87,Esc74,
Tar97,Sak99,Oza00,Tho03,Tho03b,Bel01,Blu05,Jew99,Thi00,
Sta04,Aum05,Sau01,Pal04,Cor04,Sta04b,Bec06} makes Oxygen an
attractive element for a first investigation (not to mention that
the theoretical analysis is significantly simplified in light
systems). In fact, the experimental effort on these isotopes has
spurred a number of analysis based either on the mean-field
method~\cite{Kha00b,Kha00,Kha02,Gia03,Obe05,vre05,Bec06} or
on the shell model~\cite{Uts99,Bro03,Sta04b,Vol05}. Mean-field
investigations of collective dynamics properties have up to now
exclusively concerned excitations about the lowest configuration
which always correspond to zero mass quadrupole moment. An
important result from both experimental and theoretical works is
the weakening of the \mbox{$N=20$} shell effect (both \nuc{26}{O} and
\nuc{28}{O} are not bound~\cite{Tar97,Sak99,Oza00,Tho03}) along
with evidences for a local subshell predicted sometimes at
\mbox{$N=14$}~\cite{Bec06} or at \mbox{$N=16$}~\cite{Obe05}.

For these reasons, we have chosen to investigate in a consistent
way the effect of large neutron excess on both isoscalar and
isovector quadrupole modes of the oxygen isotopes. This first
article presents our method and gives an illustration for the
nucleus \nuc{20}{O}. In order to treat small- and large-amplitude
collective motion on the same footing, we perform our analysis
within the framework of the projected generator coordinate method
(GCM)~\cite{VHB00,BBD04,BHe05} based on Hartree-Fock-Bogoliubov
(HFB) wave functions. A consistent simultaneous treatment of
isovector and isoscalar dynamics is attempted by working in a GCM
collective space in which neutron and proton deformations are
allowed to vary independently.
%
%
\section{Definition of the collective space}

Let us briefly summarize the different steps involved in the
projected configuration mixing calculation (see 
Refs.~\cite{VHB00,BBD04,BHe05} for more details).

In the GCM, the collective dynamics is constrained by the choice
of a variational space, the so-called collective space. For the
present study, this space is generated by a non-orthogonal basis
of particle-number and angular-momentum projected HFB states.

\subsection{Constrained HFB}

The HFB states are obtained from self-consistent calculations using either
a single constraint on the axial mass quadrupole moment, or a double
constraint acting separately on the axial neutron and proton quadrupole
moments. The resulting HFB states are denoted $| q \rangle$,
where $q$ labels the collective variables.
In the case of a single constraint, $q$ is identified with
the expectation value $\langle \hat{Q} \rangle$. The dynamics will
be studied along only one collective dimension, the total
deformation. The relative contributions of $\langle \hat{Q}_n \rangle$
and $\langle \hat{Q}_p\rangle$ to the total quadrupole moment
$\langle \hat{Q}_t \rangle$ are then entirely determined by the CHFB
minimization process. In the case of double contraint, the proton
$\langle \hat{Q}_p \rangle$ and neutron $\langle \hat{Q}_n\rangle$
quadrupole moments are taken as independent constraints, and the
collective variable $q$ becomes two-dimensional,
$q=\{\langle \hat{Q}_n\rangle, \langle \hat{Q}_p\rangle\}$.

As in our previous studies~\cite{VHB00,BBD04,BHe05}, the
Lipkin-Nogami (LN) prescription is used to avoid a spurious
collapse of pairing correlations in the HFB wave functions for all
values of the quadrupole moment. The HFB-LN equations are solved
with the two-basis method described in Ref.~\cite{Gal94a}. The HFB
states are chosen to be time-reversal invariant, and the
single-particle states are eigenstates of parity, $z$ signature,
and projection of isospin as in Ref.~\cite{Ter96a}. They are
represented on $1/8$ of a cubic mesh of size 11.6 fm with a
distance of 0.8 fm between mesh points.

We use the effective Skyrme interaction SLy4 in the particle-hole
channel \cite{CBH98} together with a surface-peaked
density-dependent zero-range force acting in the particle-particle
(pairing) channel. Its strength is taken equal to $-1000$ MeV
fm$^3$ in connection with two soft cutoffs at 5 MeV above and
below the Fermi energies as introduced in Ref.~\cite{BFH03}.

As a matter of fact, the one-body potentials associated with
the standard axial quadrupole operator
\begin{equation}
\label{Quadrupole}
\hat{Q}
= 2 \hat{z}^2 - \hat{x}^2 - \hat{y}^2
\end{equation}
have a numerically pathological behavior at large distances.
Indeed, depending on whether the deformation is prolate or oblate,
the constraining field decreases as $-r^{2}$ along the $z$ axis or in
the perpendicular plane, respectively. When solving the mean-field
equations on a mesh in coordinate space, the finite size of the mesh
cuts the constraint at large distances, but there always remain
attractive pockets at the boundaries of the box, which will either
distort the single-particle states around the continuum threshold,
or even bind them.
\footnote{When the CHFB
calculations are performed by expanding single-particle orbitals
on a basis such as an oscillator basis, the effective cutoff is
provided by the long-range behavior of the basis wave functions.
}
As soon as these states have a sizable occupation, as it happens in
neutron-rich systems, this will often lead to spurious results.
The large box sizes that have to be chosen in neutron-rich systems
as discussed here even amplify the problem; hence, the constraint
has to be damped at large distances from the nucleus. We achieve
this by multiplying the constraining operator with a Fermi function
as proposed in Ref.~\cite{RMR95}. We first define an equidensity
surface on which the nucleon density $\rho$ is equal to one tenth
of the maximal density $\rho_{\text{max}}$. The cutoff takes place
smoothly over a width equal to $a$ for radii exceeding the surface
by a distance $\delta r_c$.

The parameters $\delta r_c$ and $a$ have to be selected such that they
do not affect the results over the range of considered deformations.
The main role is played by the cutoff distance $\delta r_c$ which
we have set to 5.5 fm, slightly larger than the recommended value
of Ref.~\cite{RMR95}, while the width parameter was set to
\mbox{$a=0.4$} fm.
%
%
\subsection[Proj]{Particle-number and angular-momentum projected GCM}

The second step of our method is the construction of the
collective basis
\begin{equation}
\label{fctpro}
| J M q \rangle
= \frac{1}{\mathcal{N}} \,
  \hat{P}^J_{M0} \, \hat{P}^N \, \hat{P}^Z \, | q \rangle
,
\end{equation}
by means of projection operators, where $\mathcal{N} =
\langle q | \hat{P}^J_{00} \, \hat{P}^N \, \hat{P}^Z \,
| q \rangle^{1/2}$ is a normalization factor.
Although the particle numbers of the intrinsic and projected states
might be chosen different, we do not make use of this freedom and
use \mbox{$N=12$} and \mbox{$Z=8$} everywhere, and drop the
particle number indices for the sake of simple notation.

The projection operator $\hat{P}^J_{MK}$ restores the rotational
invariance broken by the intrinsic deformation of the nucleus
and provides a state with the good total angular momentum $J$.
As we consider only axial deformations with the $z$ axis as symmetry
axis, the projection of the third angular momentum component $K$ in the
intrinsic frame is equal to zero by construction. We limit our study
to values of the total angular momentum $J$ equal to zero and two.

In the next stage, we construct collective wave functions from a
linear combination of projected states
\begin{equation}
\label{fctgcm}
| J M k\rangle
= \sum_{q} f^J_{k q} | J M q\rangle
.
\end{equation}
The weights $f^J_{k,q}$ of the projected HFB states are determined
from the minimization of the total energy
\begin{equation}
\label{enegcm}
E^J_k
= \frac{\langle J M k | \hat{H} | J M k \rangle}
       {\langle J M k | J M k \rangle}
.
\end{equation}
The index $k$ has been introduced to label the several solutions
of the minimization equation which we will assign to the GCM
ground state and to excited states for each angular momentum $J$. The
variation of the energy (\mbox{$\delta_f E^J_k = 0$}) leads to the
Hill-Wheeler-Griffin equation
\begin{equation}
\label{hwequ}
\sum_{q'}
\big(   \mathcal{H}^J_{q q'}
      - E^J_k \, \mathcal{I}^J_{q q'} \big)
\, f^J_{k q'}
= 0
.
\end{equation}
Its solution requires to compute the off-diagonal matrix elements of the
energy and overlap kernels
\begin{eqnarray}
\label{kergcm}
\mathcal{H}^J_{q q'}
& = & \langle J M q | \hat{H} | J M q' \rangle
      \\
\mathcal{I}^J_{q q'}
& = & \langle J M q | J M q'\rangle
,
\end{eqnarray}
which are evaluated with the techniques described in \cite{VHB00}.
The calculation of electromagnetic transition of order $L$
requires an additional evaluation of the projected GCM kernels associated
with the electric and magnetic tensor operators $\hat{T}^{[L,M=0]}$:
\begin{equation}
\label{TLM}
\mathcal{T}^{J,[L,0]J'}_{q q'}
=  \langle J M q |  \hat{T}^{[L,0]} | J' M' q' \rangle
,
\end{equation}
where $J$, $L$ and $J'$ satisfy the triangle inequalities.

Two comments are in order. First, aside from restoring the
symmetries broken at the mean-field level, the projection also
performs a transformation from the intrinsic to the laboratory
frame. As a consequence, every state $| J=0 \, M=0 \, q\rangle$ is
\emph{spherical} in the laboratory frame of reference irrespective
of the intrinsic quadrupole deformation associated with the
collective variable $q$. Second, the coefficients $f^J_{k q}$ are
directly connected to the collective wave-function $g^J_{k q}$
which provide informations on the spreading of this wave function
over the collective space spanned by the variable(s) $q$. A set of
orthogonal collective wave functions $g^J_{k q}$ is given by
\cite{Bon90}
\begin{equation}
\label{gjkq}
g^J_{k q}
= \sum_{q'} (\mathcal{I}^J)^{1/2}_{q q'} \,
  f^J_{k q}
,
\end{equation}
where $(\mathcal{I}^J)^{1/2}$ represents the kernel whose square
convolution gives $\cal {I}^{J}$.

Numerically speaking, the GCM calculations are quite demanding.
Given the number of grid points selected on the $\{\langle
\hat{Q}_n\rangle,\,\langle \hat{Q}_p\rangle\}$ surface (about 25), the
number of gauge angles to restore the particle numbers (7 for both protons
and neutrons) and the number of Euler required for an accurate
angular-momentum projection (12), each calculation corresponds to the mixing
of about 15 thousand $N$-body wave functions. An additional
difficulty is related the non-orthogonality of the GCM basis that
generates near-zero eigenvalues of the overlap kernel
$\mathcal{I}_{q q'}$, which might lead to spurious states.
These difficulties are treated with techniques described in
Refs.~\cite{VHB00,BBD04,BHe05}.
%
%
\section[Res]{Results}

\begin{figure}[t!]
\includegraphics{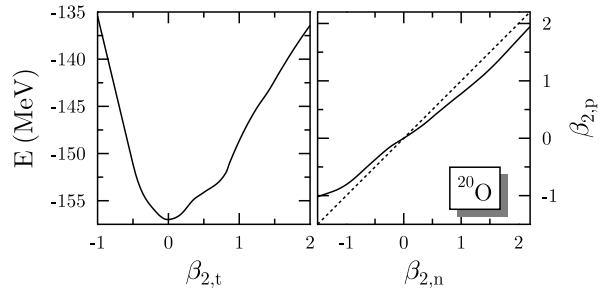}
\caption{\label{Path}
Particle-number projected energy curve obtained with a single constraint
on the mass quadrupole moment (left) and the corresponding path in the
plane of proton and neutron quadrupole deformation (right).
The dotted line in the right panel corresponds to the line of equal
proton and neutron deformation $\beta_{2,n} = \beta_{2,p}$.
}
\end{figure}

In the left panel of Fig.~\ref{Path}, we show the CHFB energy
curve obtained with a single constraint on the mass quadrupole
deformation as a function of the dimensionless quadrupole
deformation $\beta_{2,t}$
\begin{equation}
\label{eq:beta:t}
\beta_{2,t}
= \sqrt{\frac{5}{16\pi}} \; \frac{4 \pi \; Q_t}{3 R^2 A}
\end{equation}
where the radius constant is given by $R = 1.2 \; A^{1/3}$ fm.
Please note that in light nuclei and at large deformation; the
values of the $\beta_{2}$ as defined through Eq.~\ref{eq:beta:t}
might be much larger than the generating multipole deformation in
a liquid drop model. Please note also that, as in our previous
works, the mean-field energy curves that are presented are already
particle-number projected, to avoid the ambiguities related to the
Lipkin-Nogami correction to the mean-field energy. In the right
panel of Fig.~\ref{Path}, we show the path followed by the single
contraint results in the two-dimensional plane of proton and
neutron deformations
\begin{subequations}
\begin{eqnarray}
\beta_{2,n}
& = & \sqrt{\frac{5}{16\pi}} \; \frac{4 \pi \; Q_{n}}{3 R^2 N}
      \\
\beta_{2,p}
& = & \sqrt{\frac{5}{16\pi}} \; \frac{4 \pi \; Q_{p}}{3 R^2 Z}
.
\end{eqnarray}
\end{subequations}
This path corresponds to the valley that is obtained in a double-contraint
calculation. Please note that, as long as the radii of the proton and
neutron density distributions are not too different, the use of $\beta_{2,n}$
and $\beta_{2,p}$ removes most of the trivial scaling of the neutron and
proton quadrupole moments associated with different values of $N$ and $Z$.
With such coordinates, equal neutron and proton deformation of
a saturating system where the neutron density is strictly proportional
to the proton density would lie on the first bisector, represented by
the dotted line in the right panel of Fig.~\ref{Path}.
Already for not too large deformations, the path diverts substantially
from the line of equal proton and neutron deformations, indicating
that for both prolate and oblate shapes the neutrons are more deformed
than the protons.

\begin{figure}[t!]
\includegraphics{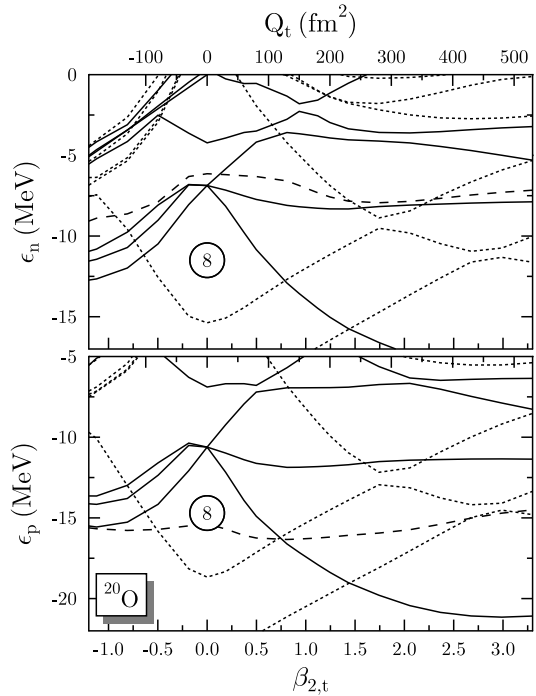}
\caption{\label{pspenergies} Neutron (top) and proton (bottom)
single-particle energies (in MeV) versus mass quadrupole
deformation (bottom horizontal scales) and mass quadrupole moment
(top horizontal scale). Solid (dotted) lines correspond to positive
(negative) parity states. The dashed lines represent the Fermi
energies of protons and neutrons, respectively.}
\end{figure}

The Nilsson diagrams of neutron and proton single-particle energies as
obtained with SLy4 for \nuc{20}{O} are shown in Fig.~\ref{pspenergies}
along the path of Fig.~\ref{Path}. Qualitatively, both are very similar.
For neutrons, however, the upward shift of the Fermi energy leads to a
particle-hole (ph) excitation at $\beta_{2,t} \approx 1.8$
($\langle \hat{Q}_t \rangle \approx 300$ fm$^2$) in which the
$\nu$ $p_{1/2}$ level is exchanged with the $\nu$ $f_{7/2}$ $m=1/2$ level.
The sequence of crossings at the Fermi surface for the protons
corresponds to a 2p-2h ($\beta\approx0.6$) and 4p-4h proton excitation
($\beta\approx3$) and is close to that obtained for \nuc{16}{O} in
Ref.~\cite{Ben03}.

\begin{figure}[t!]
\includegraphics{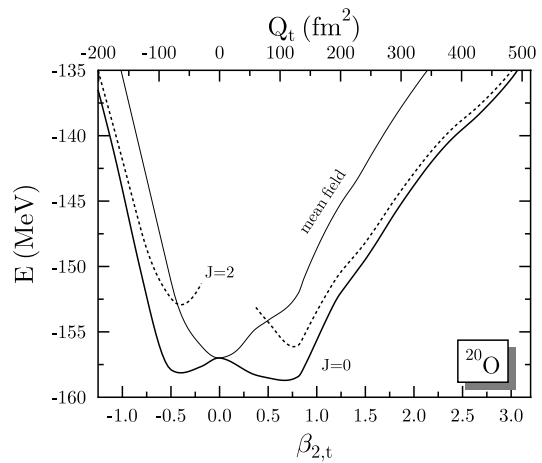}
\caption{\label{EIsoscalar}
Particle-number projected mean-field (solid curve) and  particle-number
and angular-momentum projected deformation energy curves of \nuc{20}{O}
as a function of the mass quadrupole deformation.}
\end{figure}

In Fig.~\ref{EIsoscalar} are drawn the deformation energy curves
corresponding to three different calculations. The
number-projected HFB energies are represented by a thin solid
curve. It displays a pattern typical for an oxygen isotope: a
well-defined spherical minimum and a steep rise of the energy
versus deformation, with a shoulder hinted at moderate proton
deformation. For the energy curve projected on \mbox{$J=0$}, the
behavior is markedly different. The energy does not vary much over
the range $-0.5 < \beta_{2,t} < 0.8$, or $-100$ fm$^2 < \langle
\hat{Q}_t \rangle < 200$ fm$^2$, equivalently. This reflects that
the 0p-0h configuration with respect to the spherical HFB solution
is the dominant component of the \mbox{$J=0$} wave function and
that this component is present in the mean-field wave function
over a rather large range of deformations. The part of the curve
beyond $\beta_{2,t} \simeq 0.8$, or $\langle \hat{Q}_t \rangle
\simeq 120$ fm$^2$, respectively, corresponds to the transition
associated with a 2p-2h proton excitation across the magic gap at
\mbox{$Z=8$}. These features have been studied in detail in
Ref.~\cite{Ben03} for \nuc{16}{O}. In particular, it has been
shown to explain successfully the very low excitation energy of
the first excited $0^+$ state of this isotope. By contrast, the
$2^+$ energy curve displays marked minima. The fact that for
almost all deformations but those close to sphericity the 
number-projected CHFB curve lies significantly above the $0^+$ 
and $2^+$ curves shows that components of higher angular momentum 
have a non-negligible amplitude in the intrinsic HFB states 
$| q \rangle$.

\begin{figure}[t!]
\includegraphics{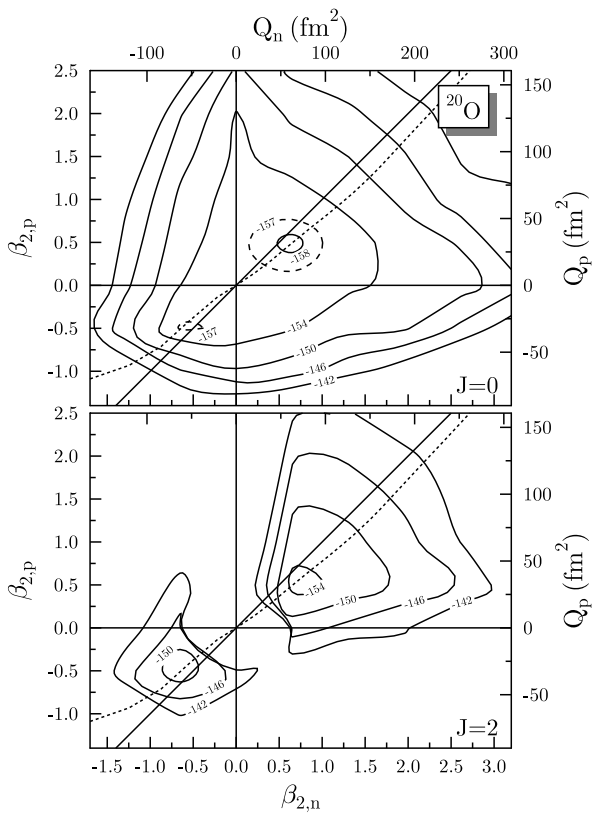}
\caption{\label{Esurface}
Contour lines of the angular-momentum and particle-particle number
projected deformation energy surface of \nuc{20}{O} (in MeV) as a
function of the neutron and proton deformation $\beta_{2,\tau}$,
$\tau=n$, $p$. The right and upper axes show the corresponding proton
and neutron quadrupole moments $Q_\tau$, $\tau=n$, $p$. The upper panel
shows the surface for $J=0$, the lower panel for $J=2$. The solid
line in both panels represents the line of equal proton and
neutron deformation $\beta_{2,p} = \beta_{2,n}$, the dotted line
the isoscalar path followed when a single constraint on the mass
quadrupole moment $\hat{Q}_t$ is applied. }
\end{figure}

\begin{figure}[t!]
\includegraphics{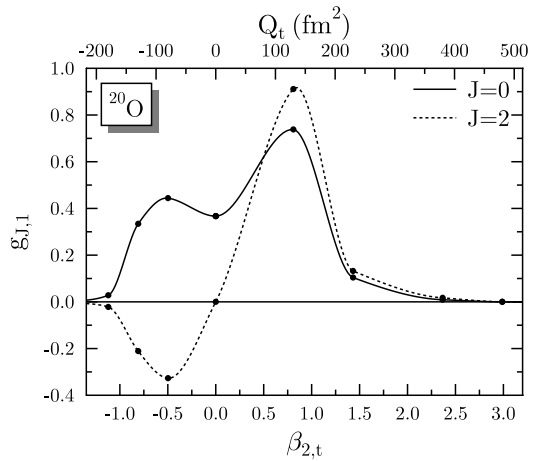}
\caption{\label{IsoscalarWF}
GCM wave functions of the $0^+$ ground
state and the lowest $J=2$ state of \nuc{20}{O} for a collective
motion restricted to one dimension, plotted versus the mass
quadrupole deformation $\beta_{2,t}$.
}
\end{figure}

In the rest of this work, we consider only neutron and proton
deformations such that the HFB excitation energy does not exceed
20~MeV. The associated zone in the $\{\langle \hat{Q}_n \rangle,
\,\langle \hat{Q}_p \rangle\}$ plane corresponds to configurations
in which the proton $1s_{1/2^+}$ and $1p_{3/2^-}$ orbitals remain
filled.

In Fig.~\ref{Esurface} are shown the \mbox{$J=0$} (top)
and \mbox{$J=2$} (bottom) projected energy surfaces. The
\mbox{$J=0$} surface displays two minima, in the quadrants where
both neutrons and protons have either prolate or oblate
deformations. One notes that the energy rises slowly along the
axis $\langle \hat{Q}_n \rangle = 0$ in contrast with the single
constraint CHFB path indicated by the dash line. A similar,
although less marked tendency exists along the $\langle \hat{Q}_p
\rangle = 0$ axis. This indicates that the $0^+$ component in
the intrinsic HFB state is mostly unchanged along these axes and
also that there already exists a relative decoupling of proton and
neutron deformations in this neutron-rich oxygen isotope.

Figure~\ref{IsoscalarWF} displays the collective GCM wave
functions $g^J_{k q}$, Eq.~(\ref{gjkq}), for the lowest $J=0$ and
$J=2$ states obtained when restricting the collective motion to
the one-dimensional path given by the dotted line in
Fig.~\ref{Esurface}. On the one hand, the \mbox{$J=0$}
ground-state wave function is localized in the vicinity of
$\langle \hat{Q}_t \rangle=0$, as expected from the potential energy
surface. On the other hand, it displays an asymmetry favoring
prolate deformations.

The representation of the two-dimensional
wave function is complicated by the fact that the discretized GCM
wave function, Eq.~(\ref{fctgcm}), is defined without a volume
element. As a consequence, the value of the collective wave
function $g^J_{kq}$ at a given point $q$ represents the wave
function times an unspecified volume element, that depends on the
distribution of the surrounding points. On the one hand, this
definition of the GCM wave function allows to optimize the set of
basis states and to remove redundant states without any
difficulty. On the other hand, any attempt to plot a GCM wave
function built from unevenly distributed points might be very
misleading as any curve used to interpolate these points to guide
the eye is, strictly speaking, meaningless. A plot of a
one-dimensional wave function $g^J_{kq}$, as given in
Fig.~\ref{IsoscalarWF}, is of course also affected, but in one
dimension one can almost always achieve a distribution of basis
states that still allows an intuitive interpretation of the 
collective wave function.
There are further issues, as the only proper metric is defined by the
overlap kernels, not the generator coordinates, which furthermore are
different for each angular momentum. These points, which affect merely
the presentation and interpretation of a GCM wave function, but not its
physical content, deserve further attention in the future.
Anyway, for the multi-dimensional wave function of \nuc{20}{O}
the points that can be chosen without having redundant
ones in the basis of projected HFB states have to be irregularly
distributed indeed.\footnote{There are in fact several near-equivalent
sets which have a very different distribution in the $\langle
\hat{Q}_n \rangle$, $\langle \hat{Q}_p \rangle$ plane.} As there is
no unambiguous way to plot a two-dimensional wave function for the
distribution of basis states we have used here, we will not
attempt to do so, but restrict the further discussion to the
observables calculated from it. However, we can state
that when adding states with a different proton-to-neutron
ratio of the deformation to one-dimensional path from the constraint on the
mass quadrupole moment, the dominating contribution to the lowest
\mbox{$J=0$} and \mbox{$J=2$} GCM states always remains the
state with $\beta_2 \approx 0.8$ located on the one-dimensional
path.

The $2^+$ energy surface (bottom part of Fig.~\ref{Esurface}) is
split into two wells in the fully prolate and oblate sectors (upper right
and lower left quadrants). In the prolate sector the energy surface
is markedly deeper. This is a good indication that the collective
dynamics will favor a prolate
deformation of the first $2^+$ state. The minimum on this surface
is shifted to the region where the neutron deformation is larger than
the proton deformation. This agrees with QRPA calculations~\cite{Kha02},
as well as with experimental evidences~\cite{Kha00}. The properties
of the first $2^+$ level are summarized in Table~\ref{Table2plus}.

In an attempt to systematically add states that cover the whole
$\beta_n$-$\beta_p$ plane to the one-dimensional calculation
depicted in Fig.~\ref{IsoscalarWF}, it turns out that the GCM
basis becomes quickly highly redundant, which is reflected by the
presence of several small eigenvalues of the norm kernel. To avoid
spurious states which would lead to large numerical inaccuracies
on energies and transition probabilities, one must avoid a very
dense and regular two-dimensional mesh. This problem is
particularly pronounced in a light nucleus as \nuc{20}{O}, in
which a SCMF wave function changes very slowly with deformation,
but becomes more marginal for nuclei beyond the $sd$ shell.
Therefore, one has to make several calculations with different
choices for the set of SCMF states, and use the variational
principle to select the mesh giving the lowest energies without
giving rise to spurious contributions. The uncertainty given in
Table~\ref{Table2plus} is a measure of the fluctuations of
the results when using a different set of points in the $\{\langle
\hat{Q}_n\rangle,\,\langle \hat{Q}_p\rangle\}$ space before entering a
zone of numerical instability. It is noteworthy that the $B(E2)$
value (and its equivalent for the neutron density distribution) fluctuate
much more in dependence of the chosen set of SCMF states than the total
energies of the first $0^+$ and $2^+$ states.

\begin{table}
\begin{tabular}{lcc}
\hline\noalign{\smallskip}
deformation space & $E$          & $B(E2\uparrow)$\\
                  & (MeV)        &  (e$^2$ fm$^4$) \\
\noalign{\smallskip}\hline\noalign{\smallskip}
 $\{\langle \hat{Q}_t\rangle\}$ & 3.1$\pm$0.1     & 80$\pm$10  \\
 $\{\langle \hat{Q}_n\rangle,\,\langle \hat{Q}_p\rangle\}$  & 3.2$\pm$0.2
    & 180$\pm$50 \\
\noalign{\smallskip}\hline\noalign{\smallskip}
Experiment       & 1.7             & 28$\pm$ 2\\
\noalign{\smallskip}\hline
\end{tabular}
\caption{\label{Table2plus}
Properties of the lowest calculated $2^+$ state in \nuc{20}{O} for
the two choices of the deformation space considered here in
comparison with experimental data taken from Ref.~\cite{Ram87}.}
\end{table}

The energy gain from the full isospin dynamics is of about 800~keV
for both the ground state and the lowest $2^+$ state, leaving the
excitation of the $2^+$ state nearly unchanged. The experimental
value is overestimated by 1.4~MeV. This discrepancy is in the
range of values observed in a global study of the properties of
the first $2^+$ states in even-even nuclei~\cite{SBB06}, performed
with a method similar to the one used. The possible origins of
this discrepancy are also discussed in Ref.~\cite{SBB06}, and for
\nuc{20}{O}, they are presumably twofold. The variational
calculation would be improved for the lowest $2^+$ state if it was
constructed from a $J_z=2$ cranked constrained set of wave
functions rather than from $J_z=0$ set as is done here. There is
also a lack of broken-pair two-quasiparticle components breaking
time reversal invariance in the variational space, which should
play a role in this semi-magic nucleus, see the discussion of
chains of heavier semi-magic nuclei in Ref.~\cite{SBB06}.

In disagreement with the global trend found by Sabbey
\etal~\cite{SBB06}, the $B(E2)$ value is in much worse agreement
with the data than the excitation energy. The effect of a larger
variational space from a double constraint is even going in the
wrong direction, leading to an overestimation of the $B(E2)$ value
by a factor 6. The value of the ratio between neutron and proton
transition matrix elements is $M_n/M_p \approx 2.1 > N/Z$. This is
less than the experimental or the QRPA values calculated with the
same force~\cite{Kha02}. It indicates that the protons follow too
closely the neutrons when deforming the nucleus, which leads to the 
too large $B(E2)$ value that we obtain. Since the QRPA value for the 
$B(E2)$ obtained with the same effective interaction is much 
larger,\footnote{It has to be
noted that the QRPA calculations of Ref.~\cite{Kha02} make
simplifying approximations for the residual interaction, which
have unclear consequences for the observables discussed here.}
this probably points out to a deficiency of our variational space
rather than to a problem due to the effective interaction. The
presence of broken-pair two-quasiparticle states breaking
time-reversal invariance in the $2^+$ wave function could reduce
the $B(E2)$ value and go into the right direction, but the present
stage of our method does not allow to test the magnitude of this
effect.
%
%
\section{Conclusions}
We have started an investigation of the large-amplitude quadrupole
dynamics in neutron-rich nuclei taking into account the full
isospin space. Among our initial motivation was the search for
stable configurations in which neutron and proton display
different static deformation. Our results indicate that it is
indeed the case when the usual dimensionless deformations
$\beta_{2,n}$ and $\beta_{2,p}$ are used to remove the trivial
scaling of the quadrupole moments with neutron and proton number.

We have also looked for signs of a dynamical isospin instability
which would show up in specific spectroscopic properties. At a
qualitative level, we observe the importance of allowing
independent proton and neutron deformation: although the main
features of the collective wave functions are still the same as
those determined by the one-dimensional dynamics, the full wave
function displays a significant extension on both sides of this
path. Similarly, there is a clear influence on the transition
moments. Unfortunately, the agreement with the experimental data
is not satisfactory for this light nucleus. Our model would
probably be improved by enlarging the variational space for the
$2^+$ state with the inclusion of states breaking time-reversal
invariance. The developments which will allow this inclusion are
still underway.
%
%
\section*{Acknowledgments}

Part of the work by M.~B.\ was performed within the framework of
L'Espace de Structure Nucl{\'e}aire Th{\'e}orique (ESNT).
This work has been partly supported by the PAI P5-07 of
the Belgian Science Policy Office.

\end{document}